# $T_c$ is insensitive to magnetic interactions in high-$T_c$ superconductors

B. Mallett[1†], G.V.M. Williams[1], A.B. Kaiser[1], E. Gilioli[2], F. Licci[2], T. Wolf[3], J.L. Tallon[4]

[1]MacDiarmid Institute, School of Physical and Chemical Sciences, Victoria University of Wellington, P.O. Box 600, Wellington 6140, New Zealand
[2]IMEM-CNR, Institute of Materials for Electronics and Magnetism, 43124 Parma, Italy
[3]Karlsruhe Institute of Technology, Inst. Solid State Physics, P.O. Box 3640, D-76021 Karlsruhe, Germany
[4]MacDiarmid Institute, Industrial Research Limited, P.O. Box 31310, Lower Hutt, New Zealand.

**A quarter of a century after their discovery the mechanism that pairs carriers in the cuprate high-$T_c$ superconductors (HTS) still remains uncertain. Despite this the general consensus is that it is probably magnetic in origin [1] so that the energy scale for the pairing boson is governed by $J$, the antiferromagnetic exchange interaction. Recent studies using resonant inelastic X-ray scattering strongly support these ideas [2]. Here as a further test we vary $J$ (as measured by two-magnon Raman scattering) by more than 60% by changing ion sizes in the model HTS system $LnA_2Cu_3O_{7-\delta}$ where A=(Ba,Sr) and Ln=(La, Nd, Sm, Eu, Gd, Dy, Yb, Lu). Such changes are often referred to as ``internal'' pressure. Surprisingly, we find $T_c^{max}$ anticorrelates with $J$ where *internal* pressure is the implicit variable. This is the opposite to the effect of *external* pressure and suggests that $J$ is not the dominant energy scale governing $T_c^{max}$.**

In this work we are motivated by a paradox. External pressure increases the $T_c^{max}$ of optimally doped cuprates [3] whereas internal pressure, as induced by isovalent ion substitution, decreases $T_c^{max}$ [4, 5]. Fig. 1(a) shows $T_c^{max}$ plotted versus the composite bond valence sum parameter, $V_+ = 6 - V_{Cu} - V_{O2} - V_{O3}$, as reproduced from reference [4] (green squares). Here $V_{Cu}$, $V_{O2}$ and $V_{O3}$ are the planar copper and oxygen BVS parameters and the plot reveals a remarkable correlation of $T_c^{max}$ across single-, two- and three-layer cuprates, where for the latter $V_+$ was calculated for the inner of the three layers. $V_+$ was introduced as a measure of charge distribution between the Cu and O atomic orbitals but is also a measure of in-plane stress. Evidently stretching the $CuO_2$ plane is effective in increasing $T_c^{max}$. $V_+$ is however a compound measure and also e.g. reflects displacement of the apical oxygen away from the Cu atoms, as occurs across the title compounds.

To this old plot we add new data for the title compounds, $LnA_2Cu_3O_{7-\delta}$, as Ln is varied across the lanthanide series and, in the case of Ln = Y, A = $Ba_{2-x}Sr_x$ for $x$ = 0, 0.5, 1, 1.25 and 2. We use the Rietveld structural refinements of Guillaume *et al*. [6], Licci *et al*. [7] and Gilioli *et al*.[8] and calculate $V_+$ in the same way as previously [4]. The new data are shown by crosses in Fig. 1(a) and in more detail in Fig. 1(b). Notably, the global correlation seen in Fig. 1(a) is preserved across the title system. Moreover we show in Fig. 1(c) a plot of $T_c^{max}$ versus cell volume and while $T_c^{max}$ increases with lattice expansion the slopes differ depending on whether Ln or A is altered. This same change in slope is also observed when $T_c^{max}$ is plotted versus the basal cell area $a \times b$ or the Cu(2)-O(2,3) bond length (which is half the superexchange path). There is clearly some additional feature that the $V_+$ correlation incorporates as the slope remains essentially the same across all cuprates in general and for the title system irrespective of whether A is altered or Ln.

Using the bulk compressibility it is straightforward to convert these volume changes to an effective internal pressure and Fig. 1 thus summarises a general feature of the cuprates, namely that *internal pressure* decreases $T_c^{max}$. In contradistinction, as already noted, it is well known that external pressure increases $T_c^{max}$ [3].

What then is the salient difference between internal and external pressure on $T_c^{max}$? The question may be posed within a weak-coupling BCS framework, which recent work has shown can describe the cuprates once the competing pseudogap phase and superconducting fluctuations are considered [9]. Specifically we take the weak-coupling *d*-wave relation:

$$k_B T_c^{max} = 0.935\ \hbar\omega_B\ \exp[-1/(N(E_F)V)] \quad (1)$$

where $T_c^{mf}$ is the mean-field superconducting transition temperature, $\omega_B$ is the pairing boson energy scale, $N(E_F)$ is the density of states (DOS) at the Fermi-level and $V$ the pairing potential.

In the underdoped regime the DOS is progressively depleted by the opening of the pseudogap (with energy scale $E_g$), whereas in the overdoped regime it is enhanced by the proximity of the van Hove singularity (vHs). It is therefore our long-term goal to study the comparative effects of internal and external pressure on the key variables, $T_c^{max}$, $\omega_B$, $V$, $N(E_F)$, $E_g$ and $E_F - E_{vHs}$. Here we focus on $\omega_B$ which if the pairing boson is a paramagnon is governed by $J$, the energy for antiferromagnetic superexchange interaction between neighbouring Cu(2) sites.

We measured the magnitude of $J$ in single crystals of LnBa$_2$Cu$_3$O$_6$ using $B_{1g}$ two-magnon Raman scattering. The technique is illustrated in Fig. 2(a). The two-magnon peak maximum occurs at $\omega_{max} = 3.2J = 2.8J_{eff}$ [10]. It has previously been found that $J$ is quite strongly doping dependent, in the case of Bi$_2$Sr$_2$Ca$_{1-x}$Y$_x$O$_{8-y}$ falling more or less linearly with doping from 125 meV in the undoped insulator to 45 meV when $p = 0.20$ holes/Cu in the lightly overdoped region [11]. Despite this strong doping dependence we focus on the undoped insulator (O$_6$) because this is the only reproducible doping state for intercomparison of the different lanthanides. And perhaps more importantly, we wish to avoid pressure-induced charge transfer which can only be assured when the CuO chains are fully deoxygenated.

The normalized raw data is shown in Fig. 2(b) as a false-colour intensity plot in the frequency-versus-ion-size plane. Intensity values were linearly interpolated between the measured Ln123 spectra, which are themselves shown in the inset to Fig. 2(c). Clearly, as the Ln ion size decreases the $B_{1g}$ two-magnon peak shifts to higher frequency, indicating that $J$ increases. Smaller Ln ion sizes [12] result in shorter in-plane bond lengths [6] and larger effective internal pressures on the CuO$_2$ layers.

The relative shift in effective internal pressure, $\Delta P_{eff}$, may be estimated from the change in unit cell volume, $\Delta V$, using $\Delta P_{eff} = -K.\Delta V/V_0$ where $K = -0.65\times10^{-3}$ GPa.nm$^{-3}$ is the bulk compressibility [13]. We plot $\omega_{max}$ against $\Delta P_{eff}$ in Fig.2(c). The observed positive gradient is expected due to increased exchange interaction arising from increased overlap between Cu(2) *d* and O(2,3) *p* orbitals.

We can quantify the shift in $J$ with Ln size using an area Grüneisen parameter, $\gamma_A = -\ln(J/J_0)/\ln(A/A_0)$
where $A$ is the basal area, $a\times b$, of the unit cell. The area dependence of $J$ is plotted in Fig. 3(a), where $A$ is calculated from Guillaume *et al.* [6]. We find $\gamma_A = 3.0 \pm 0.6$ where the large uncertainty reflects the fact that Yb seems to be an outlier.

The figure also compares this *internal pressure* effect with the effect of *external pressure* on $J$ in the related system La$_2$CuO$_4$ [14]. This is shown by the blue diamonds in Fig. 3(a) which range from ambient pressure to 10 GPa, as annotated. Remarkably, the dependence of $J$ on $A$

(= $a \times b$) is preserved across the entire range including two quite structurally disparate cuprates. Collectively these cover a 50% increase in the magnitude of $J$ arising from simple structural compression. From these measurements we calculate $\gamma_A = 3.1 \pm 0.1$ for $La_2CuO_4$ in good agreement with what we find for Ln123. We conclude that external pressure has a similar effect to internal pressure on $J$ – contrary to their known effects on $T_c^{max}$.

The only published data for two-magnon scattering in Y123 under external pressure is for $YBa_2Cu_3O_{6.2}$ [15] and this is plotted by the green diamonds in the inset to Fig. 3(a) along with our data for Ln123. The residual 0.2 oxygens in the chain layer introduce uncertainties around possible pressure-induced charge transfer and we are therefore cautious in the use of this data. Nonetheless, they reveal a trend which is fully consistent with the ion-size effect for Ln123, again confirming that internal and external pressure have similar effects on the magnitude of the exchange interaction.

Turning now to our main result we plot in Fig. 3(b) $T_c^{max}$ versus $J$ for the Ln123 single-crystal series. We have used $T_c^{max}$ = 98.5 K for La123 [16] and $T_c^{max}$ =96 K for Nd123 [17] as these are the highest reported values of $T_c^{max}$ in these compounds (where Ln occupation of the Ba site is minimised). Surprisingly and contrary to all expectation from Eq. 1, $T_c^{max}$ *anticorrelates* with $J$ when ion-size is the implicit variable.

Guided by Fig. 1, and in an attempt to push out to higher $J$ values, we repeated the Raman measurements on a *c*-axis aligned thin film of $YBa_{1.5}Sr_{0.5}Cu_3O_6$ and on individual grains of polycrystalline $YSr_2Cu_3O_6$, prepared under high-pressure/high-temperature synthesis [8]. The same anticorrelation between $T_c^{max}$ and $J$ is preserved, and in this case now out to a more than 60% increase in the value of $J$.

This result may appear to be at odds with other Raman studies where it is reported that, with increasing doping, the $B_{1g}$ gap parameter observed in low-frequency Raman spectroscopy is proportional to $J$ [11]. But it is important to note that the $B_{1g}$ gap is not the superconducting order parameter, $\Delta_0$. $B_{1g}$ probes anti-nodal regions of the Fermi-surface so will contain contributions from both the pseudogap, $E_g$, and $\Delta_0$, and at low doping is completely dominated by $E_g$. Consequently, this reported correlation of the $B_{1g}$ gap with $J$ just establishes a direct correlation between $E_g$ and $J$ as has already been inferred from specific heat [18] and inelastic neutron scattering [19] experiments. It is the low-frequency $B_{2g}$ gap feature which probes $\Delta_0$.

Thus we conclude that $T_c^{max}$ anticorrelates with $J$ where internal pressure is the implicit variable. This is not the case for external pressure – both $T_c^{max}$ and $J$ increase with external pressure, so $T_c^{max}$ must *correlate* with $J$ when external pressure is the implicit variable.

Our observations suggest three possibilities; (i) $T_c$ is unrelated to $J$. Though we feel this is unlikely it would mean that the pseudogap alone is governed by $J$ and is enhanced by both external and internal pressure. (ii) The doping dependence of $J$ [11] is radically different across the Ln123 series and by optimal doping their trend with ion size has reversed. We are currently investigating this possibility. Or (iii), other energy scales also vary across the Ln123 series and have a much larger effect on $T_c^{max}$ than $J$ or $\omega_B$. These might include: $V$, $E_g$, $E_F - E_{vHs}$, interplanar Josephson coupling, scattering rates ($\Gamma$), or condensation energy (governing fluctuations [9]). Of these we consider it most likely that increasing ion size shifts the vHs closer to the Fermi level by altering the relative magnitudes of the $t$, $t'$ and $t''$ hopping parameters [20]. The combination of an exponentially-strong DOS dependence that is also divergent can have profound effects on the systematic evolution of $T_c$ with ion size.

In conclusion we have shown the anti-ferromagnetic superexchange energy, $J$, decreases as the Ln ion size in Ln123 increases. Furthermore, the effect of internal pressure on $J$, from Ln substitution, is quantitatively similar to the effect of external pressure on $J$ across a very wide range of compression, encompassing an increase of $J$ by 50%. This is to be contrasted with the opposite effects of internal and external pressure on $T_c^{max}$. We find that $T_c^{max}$ anticorrelates with $J$ when ion size is the implicit variable which suggests some energy scale other than short-range anti-ferromagnetic interactions or paramagnons has a more dominant effect on $T_c^{max}$.

**Method.** High-quality $LnBa_2Cu_3O_{7-\delta}$ single crystals were flux grown in Y-stabilized zirconia crucibles under reduced oxygen atmosphere, where necessary, to avoid substitution of Ln ions on the Ba site. The crystals were annealed for three days in flowing Ar at 600°C to remove oxygen from the chains then quenched to room temperature while still under Ar gas. We expect an oxygen deficiency of $\delta \geq 0.97$ from these annealing conditions and the measured mass change. Fully de-oxygenated chains are inert and this should eliminate any internal pressure-induced charge transfer that could otherwise occur between chain and $CuO_2$ layers. Moreover, even if there is some residual oxygen in the chains the doping state remains zero provided the oxygens remain isolated and chain segments do not start to form [21, 22].

Well-oriented films of $Y(Ba_{1-x}Sr_x)_2Cu_3O_{7-\delta}$ were synthesised by metallo-organic deposition (as is used for the preparation of 2G superconducting tapes) on RaBITS™ textured Ni/W substrates with epitaxial buffer layers of $Y_2O_3/YSZ/CeO_2$. Polycrystalline samples of $YSr_2Cu_3O_{7-\delta}$ were prepared at 3 GPa and 1050°C using a multi-anvil apparatus with $KClO_3$ as an oxidant [8].

Raman measurements were made at ambient temperature using a LabRam confocal microscope Raman spectrometer in back-scattering $B_{1g}$ geometry. For all samples except the polycrystalline $YSr_2Cu_3O_6$ the $B_{2g}$, $A_{1g}+B_{2g}$ and $A_{1g}+B_{1g}$ scattering geometries were used to check we were indeed measuring two-magnon scattering. The 514.5 nm line from an Ar ion laser with power $\leq 1\,mW$ was used and focused to a spot of size ~1μm. A 300 lines/mm diffraction grating was used to capture the two-magnon peak in a single frame.

**Acknowledgements.** We thank the Marsden Fund and the MacDiarmid Institute for financial assistance.

† Corresponding author. E-mail: b.mallett@irl.cri.nz

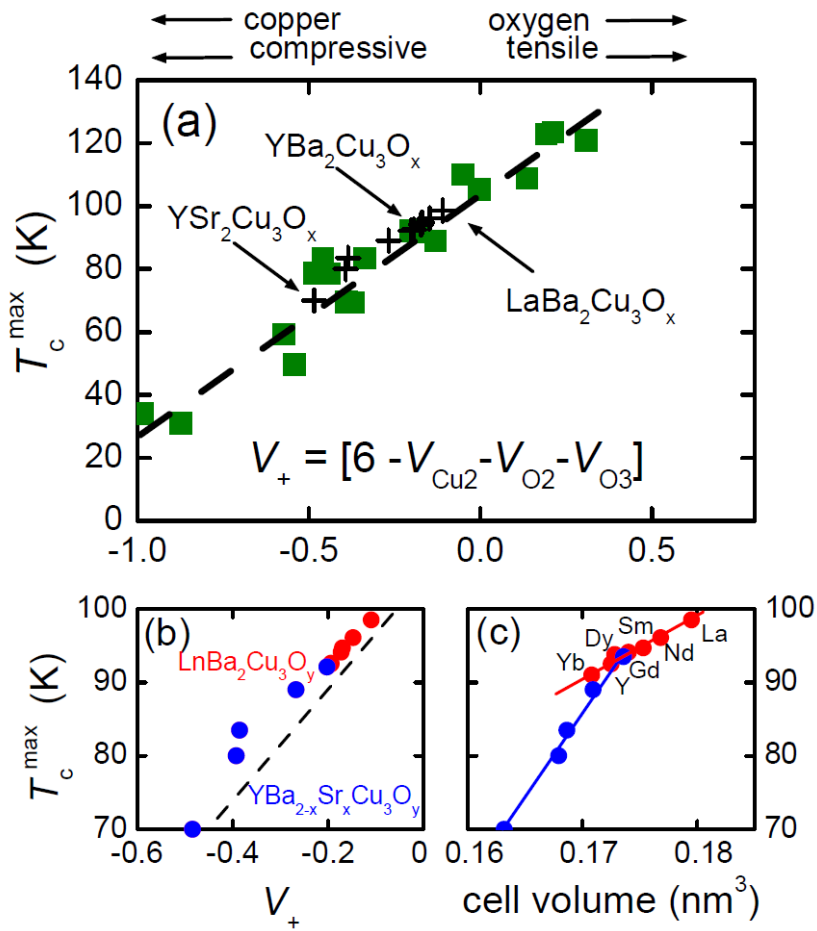

Figure 1 (a) $T_c$ at optimal doping, $T_c^{max}$, plotted as a function of the compound bond valence sum parameter $V_+ = 6 - V_{Cu2} - V_{O2} - V_{O3}$. Green squares: as previously reported [4]; crosses: for the title materials LnA$_2$Cu$_3$O$_7$. (b) shows $T_c^{max}$ versus $V_+$ for LnBa$_2$Cu$_3$O$_7$ (red circles) and for Y(Ba,Sr)$_2$Cu$_3$O$_7$ (blue circles), confirming the general correlation also holds for the title compounds. (c) shows $T_c^{max}$ versus cell volume for LnBa$_2$Cu$_3$O$_7$ (red circles) and for Y(Ba,Sr)$_2$Cu$_3$O$_7$ (blue circles), While $T_c^{max}$ is raised by expanding the lattice the effect is different depending on whether executed on the Ln site or the Ba site.

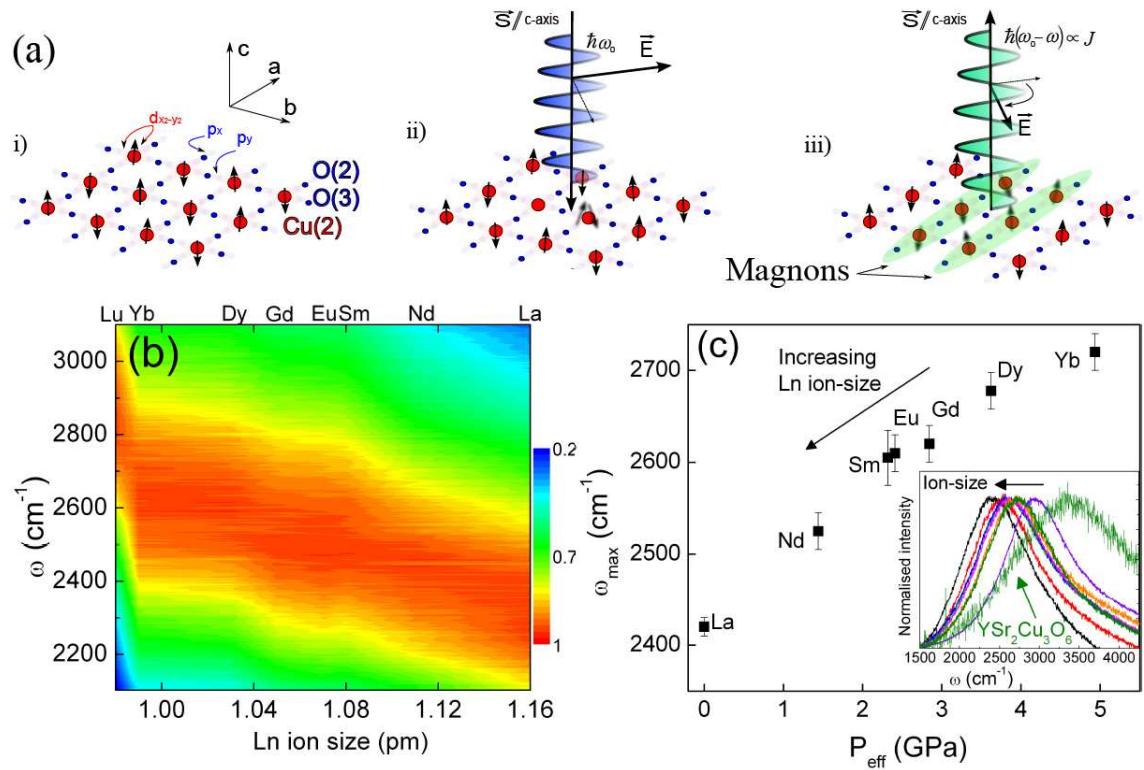

Figure 2 (a) Schematic diagram of the simplified $B_{1g}$ two magnon scattering process; (i) the undoped $CuO_2$ layer has long range anti-ferromagnetic order. (ii) An absorbed photon excites a spin to doubly occupy an adjacent site. (iii) Relaxation occurs by the opposite spin filling the empty site accompanied by the emission of red-shifted photon. Two magnons are created from this two-spin flip process with a rotation in polarisation of Raman scattered photon and energy loss proportional to $J$.

(b) Normalized Raman spectra as a function of Ln ion-size with the intensity of $B_{1g}$ two-magnon peak of $LnBa_2Cu_3O_6$ (Ln123) in false colour. The spectral shift to higher frequencies as the ion-size decreases can be clearly seen. A linear interpolation between measured Ln123 spectra was applied.

(c) The two-magnon peak maximum, $\omega_{max}$, for each Ln123 sample plotted against the effective internal pressure referenced to La123, calculated from $P_{eff} = K(V - V_0)$ where $K = -0.65 \times 10^{-3}$ GPa.nm$^{-3}$ is the bulk compressibility per unit volume [13]. Inset shows normalized Raman spectra of (b) for Ln123 single crystals and polycrystalline $YSr_2Cu_3O_6$.

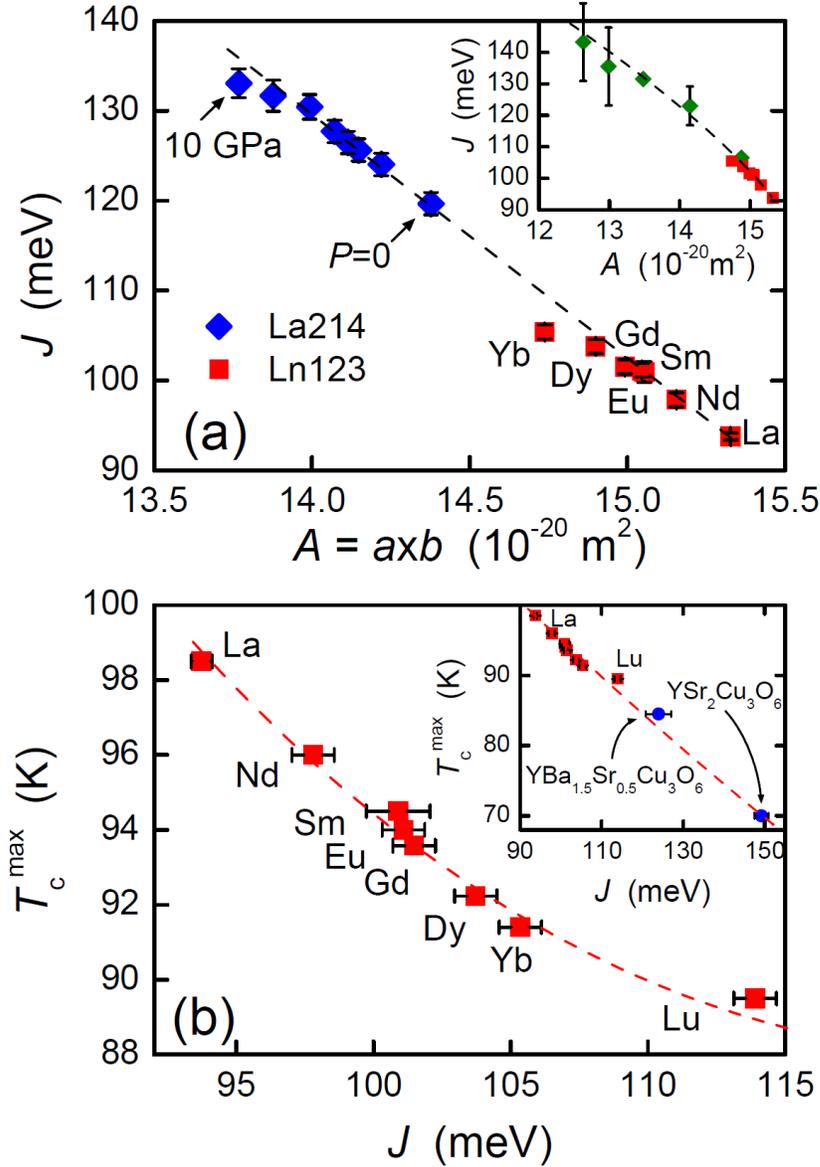

Figure 3(a) The basal area dependence of $J$ determined from our two magnon scattering studies on LnBa$_2$Cu$_3$O$_6$ (red squares) where internal pressure (or ion size) is the implicit variable. This is compared with the same plot for La$_2$CuO$_4$ where external pressure (0 to 10 GPa) is the implicit variable [14]. Remarkably a single behaviour for $J(A)$ is preserved across a 50% change in $J$ irrespective of whether the effective pressure is internal or external. We find $\gamma_A = 3.0 \pm 0.6$ for Ln123, $\gamma_A = 3.1 \pm 0.1$ for La214. Inset: shows $J$ versus $A$ for pressure-dependent two-magnon scattering data for YBa$_2$Cu$_3$O$_{6.2}$ to 80 GPa [15] (green diamonds) compared with our ion-size-dependent data for LnBa$_2$Cu$_3$O$_6$ (red squares). Again external and internal pressures appear to have quantitatively similar effects on $J$. The dashed line is a guide to the eye.

(b) $T_c^{max}$ is plotted versus $J$ for our single crystals of LnBa$_2$Cu$_3$O$_6$ revealing a systematic *anticorrelation* of $T_c$ with magnetic exchange interactions. Inset: extends this plot to include $T_c^{max}$ versus $J$ for Y(Ba$_{1-x}$Sr$_x$)$_2$Cu$_3$O$_6$ (blue circles) for $x = 0.25$ and 1.0. Remarkably this anticorrelation extends over a more than 60% change in $J$.